\documentclass[aps,prd,preprint,superscriptaddress,showpacs,floatfix]{revtex4}

\usepackage{graphicx}
\usepackage{longtable}

\begin{document}

\title{Pion-nucleon Sigma Term in the Global Color Model of QCD}

\author{Lei Chang}
\affiliation{Department of Physics, Peking University, Beijing
100871, China}

\author{Yu-xin Liu}
\email[corresponding author]{} \affiliation{Department of Physics,
Peking University, Beijing 100871, China} \affiliation{The Key
Laboratory of Heavy Ion Physics, Ministry of Education,Beijing
100871, China }  \affiliation{Center of Theoretical Nuclear
Physics, National Laboratory of Heavy Ion Accelerator, Lanzhou
730000, China} \affiliation{CCAST(World Lab.), P. O. Box 8730,
Beijing 100080, China}

\author{Hua Guo}
\affiliation{Department of Technical Physics, Peking University,
Beijing 100871, China} \affiliation{The Key Laboratory of Heavy
Ion Physics, Ministry of Education,Beijing 100871, China }
\affiliation{Center of Theoretical Nuclear Physics, National
Laboratory of Heavy Ion Accelerator, Lanzhou 730000, China}

\date{\today}

\begin{abstract}
We study the pion-nucleon sigma term in vacuum and in nuclear matter
in the framework of global color model of QCD. With the effective
gluon propagator being taken as the $\delta$-function in momentum
space of Munczek-Nomirovsky model, we estimate that the sigma term
at chiral limit in the vacuum is 9/2 times the current quark mass
and it decreases with the nuclear matter density. With the presently
obtained in-medium pion-nucleon sigma term, we study the in-medium
chiral quark condensate and obtain a reasonable variation behavior
against the nuclear matter density.

\end{abstract}


\maketitle

\bigskip

\newpage

\section{Introduction}

The pion-nucleon sigma term $\sigma_{\pi N}$ is of fundamental
importance for understanding the chiral symmetry breaking effect
in nucleon\cite{Reya74,Jaffe80,Gasser81,Sainio02}, the central
nuclear force\cite{MPR00,Robi01} and the mass decomposition of
nucleon\cite{Ji95,Procura04}. Because it is related to the quark
and gluon condensates in nuclear
matter\cite{TMF91,CFG92,Birse94,Fuji95,Dey95,Brock96,Lutz98,Brown02,Dru013},
the pion-nucleon sigma term is regarded to play an important role
in the process of chiral symmetry restoration in nuclear matter.
Recent researches show that the $\sigma_{\pi N}$ is also important
in searching for the Higgs boson\cite{Cheng88}, supersymmetric
particles and dark matter\cite{BDFS00,CCN02}. Then the
pion-nucleon sigma term has been studied in chiral perturbation
theory\cite{BKM936,Borasoy9679,OW99}, lattice
QCD\cite{Fukugita95,DLL96,Guesken99,Thomas004,Procura04}, various
chiral models\cite{BRS88,DPP89,GLS91,BM92,LTTW00,LGFD01,Schw041}
and other models\cite{Peng02,Peng05,Hite05}. As for the value of
the pion-nucleon sigma term, different researches show that it can
be as small as ($18\pm5$)~MeV\cite{Guesken99}, and as large as
(88-90)~MeV\cite{Schw042}. Generally it used to be regarded as
45~MeV\cite{GLS91,LGFD01}, and recently suggested to be
(50-80)~MeV\cite{OK02,Schw041,Schw042,Hite05}. Since experiment
can not measure the $\sigma_{\pi N}$ directly and theoretical
results are model dependent, the value of the sigma term has been
and still is a puzzle\cite{Schw041}. On the other hand, when
taking the sigma term as an ingredient to determine the quark and
gluon condensates in nuclear matter, it was usually taken as a
constant independent of the nuclear matter
density\cite{TMF91,CFG92,Birse94,Fuji95,Dey95,Brock96,Lutz98,Brown02,Dru013}.
In such a sense, the nuclear matter density dependence of the
pion-nucleon sigma term has not yet been studied.

It has been shown that the global color model
(GCM)\cite{CR858,FT91,FT92,JF97,Tan97,LLZZ98} is a quite
successful effective field theory model of QCD in describing
hadron properties and quark condensate\cite{Meis97,KM98,Zong} in
free space(i.e., at temperature $T=0$, chemical potential $\mu
=0$). Meanwhile the bag constant, the radius and the mass of a
nucleon in nuclear matter and the quark condensate in nuclear
matter can also be evaluated in a consistent
way\cite{Liu,MRS98,BPRS98,LGG03} in the GCM. With the global color
symmetry model at zero and finite chemical potential $\mu$, we
will study the pion-nucleon sigma term in free space and in
nuclear matter in this paper. As an application of the sigma term
in nuclear matter, we will also discuss the density dependence of
the chiral quark condensate in nuclear matter.

The paper is organized as follows. In section II, we describe
briefly the framework of the global color symmetry model. In section
III, we give our model and result of the pion-nucleon sigma term. In
section IV, we apply the obtained pion-nucleon sigma term to
evaluate the nucleon matter density dependence of the chiral quark
condensate. Finally, we summarize our work and give a brief remark
in section V.

\section{Brief description of the Global Color Model of QCD}

We start from the action of the global color model of QCD in
Euclidean space
\begin{eqnarray}
S_{GCM}[\bar{q},q]& = & \int d^{4} x
\left\{\bar{q}(x)[\gamma\cdot\partial_{x}+m]q(x)\right\}  \nonumber
\\ & & +\frac{1}{2}\int d^{4}x d^{4} y\left[j^{a}_{\mu}(x)g^2_{s}
D^{ab}_{\mu\nu}(x-y)j^{b}_{\nu}(y)\right] \, ,
\end{eqnarray}
where $m$ is current quark mass,
$j^{a}_{\mu}(x)=\bar{q}(x)\gamma_{\mu}\frac{\lambda^{a}_{c}}{2}q(x)$
denotes the color octet vector current and
$g^2_{s}D^{ab}_{\mu\nu}(x-y)$ is the effective gluon propagator.
Here we would like to diagonal the gluon propagator in the color
matrix and choose the Landau gauge, i.e., take the effective gluon
propagator as
\begin{equation}
g_{s}^2 D^{ab}_{\mu\nu}(x)=\delta_{ab}\int\frac{d^{4}k}{(2\pi)^4}
t_{\mu\nu} (k)\mathcal{G}(k^2)e^{ik\cdot x} \, ,
\end{equation}
where $ t_{\mu\nu}(k) = \delta_{\mu\nu}-\frac{k_{\mu}k_{\nu}}{k^{2}}$
and $\mathcal{G}(k^2)$ is the effective interaction relating to
the gluon vacuum polarization introduced usually as a model input.

Introducing the auxiliary bilocal fields $B^{\theta}(x,y)$ and
applying the standard bosonization procedure\cite{CR858}, we can
give the partition function at mean field approximation as
\begin{equation}
{\cal{Z}}=\int{\cal{D}} B^{\theta} e^{-S_{eff}[B^{\theta}]} \, .
\end{equation}
One can then identify the auxiliary field that minimizes the
effective action as $B^{\theta}_{0}$ (usually referred to as the
vacuum configuration). Expansion in filed fluctuations about the
vacuum configuration would generate the propagating bosons (mesons
etc). Meanwhile, the vacuum configuration produces the rainbow
approximation for the quark self-energy giving the rainbow
Dyson-Schwinger equation
\begin{equation}
G^{-1}(p)=i\gamma\cdot p+m+\frac{4}{3}\int\frac{d^{4}q}{(2\pi)^{4}}
t_{\mu\nu}(p-q)\mathcal{G}((p-q)^2)  \gamma_{\mu}G(q)\gamma_{\nu}  \, ,
\end{equation}
where $G(q)$ is the   % dressed   % $\cdots \cdots$ }
%\begin{equation}
%G^{-1}(p)=i\gamma\cdot p+m+\frac{4}{3}\int\frac{d^{4}q}{(2\pi^2)}
%D_{\mu\nu}(p-q)\gamma_{\mu}G(q)\gamma_{\nu} \, ,
%\end{equation}
%where $D_{\mu \nu}$ is the effective gluon propagator, $G(p)$ is the
dressed quark propagator which can be decomposed as
\begin{equation}
G^{-1}(q)=i\gamma\cdot q A(q^2)+B(q^2) \, ,
\end{equation}
with  $A(q^2)$, $B(q^2)$ being scalar self-energy function and
$B(q^2) = B^{\theta}_{0}$. Since the rainbow approximation of the
Dyson-Schwinger equation determines the vacuum configuration and the
quantum fluctuations about this configuration add corrections to the
rainbow approximation, the GCM is a quite sophisticated and
practical effective field theory model of QCD.

It is remarkable that the $B(p^2)$ deduced within an effective
interaction from the quark equation in the chiral
limit($m\rightarrow 0$) has two qualitatively distinct solutions.
One is the Wigner solution, $B(p^2)\equiv 0$, that characterizes the
phase in which chiral symmetry is not broken and the dressed quarks
are not confined. Another is the Nambu solution with nonzero
$B(p^{2})$, which describes the phase where chiral symmetry is
spontaneously broken and prevents quarks from involving mass pole
bellow about 1-2 GeV\cite{RW94}. It is also necessary to mention
that the Wigner solution is always possible, but the Nambu solution
is only available if the coupling is strong enough at the infrared
region\cite{RW94}.

For studying the pion-nucleon sigma term, we should know the
derivative of the constituent quark mass against the current quark
mass. It is at first necessary to start from the current mass
dependent quark propagator $G(p)$. Defining the derivative of the
inverse of the quark propagator with respect to the current quark
mass as
\begin{equation}
\Gamma(p^2)=\frac{\partial G^{-1}(p)}{\partial m} \, ,
\end{equation}
one can easily prove that the $\Gamma(p^2)$ satisfies the equation
\begin{equation}
\Gamma(p)=1-\frac{4}{3}\int\frac{d^{4}q}{(2\pi)^{4}} {t}_{\mu\nu}(p-q)
\mathcal{G}((p-q)^2) \gamma_{\mu}G(q)\Gamma(q)G(q) \gamma_{\nu} \, .
\end{equation}
Analyzing the Lorentz structure, one can decompose the
$\Gamma(p^2)$ function as
\begin{equation}
\Gamma(p^2)=i\gamma\cdot p C(p^2)+D(p^2)  \, .
\end{equation}
This equation together with the quark equation has been used to
study the nonperturbative mass-independent renormalization within
the Dyson-Schwinger equation formalism\cite{Toki01}. Here it
should be noted that not only the scalar part of the quark
propagator $B(p^2)$ depends on the current quark mass but also the
vector part $A(p^2)$. At the same time, it is useful to
simultaneously discuss the spontaneous chiral symmetry breaking
and the explicit symmetry breaking.

\section{Pion-nucleon Sigma Term in the Global Color Model of QCD}

To understand the spontaneous chiral symmetry breaking, one usually
take the chiral susceptibility as a manifestation. The chiral
susceptibility is defined as\cite{Holl01}
\begin{equation}
\chi(p^2)=\frac{\partial M(p^2)}{\partial m}  \, ,
\end{equation}
where the quark mass function $M=B/A$ can be fixed by implementing
the quark equation. With the quark propagator and its derivation
of the current quark mass we can give explicitly the chiral
susceptibility as
\begin{equation}
\chi(p^2)=\frac{D(p^2)A(p^2)-B(p^2)C(p^2)}{A^{2}(p^2)} \,  .
\end{equation}

The pion-nucleon sigma term is related to the nucleon mass $M_{N}$
by means of the Hellmann-Feynman theorem
\begin{equation}
\sigma_{\pi N} = m \frac{\partial M_{N}}{\partial m} \, .
\end{equation}
The nucleon will be treated as three non-interacting constituent
quarks so as to emphasize the dominant characteristics and permit
simple estimates in spirit of the chiral quark soliton
model\cite{Schw041,CR858} in the isospin symmetry limit. We have
then $M_{N} = 3 M_{c}$. The quark constituent mass will be evaluated
by the so-called Euclidean constituent mass which is the value of
mass function matching the momentum scale and reads $M_{c}=
M(p^{2}=M_{c}^{2})$. With the chiral susceptibility, the
pion-nucleon sigma term can then be expressed as
\begin{equation}
\sigma_{\pi N} = 3 m \chi(p^{2}=M_{c}^2)_{m\rightarrow 0}  \, .
\end{equation}
Here we have taken the chiral limit for the chiral susceptibility on
the ``quark mass shell". We should note that quark confinement
entails only that there is no ``pole mass". The constituent quark
mass $M_{c}$ may then be estimated in other ways. One of the other
choices could be the value of the quark dynamical mass function at
$p^{2} =0$, that is $\tilde{M_{c}}= M (p^{2} = 0) $. We will also
consider this choice.

In the following practical calculation, we take the
Munczek-Nomirovsky model\cite{MN83} of the effective gluon
propagator
\begin{equation}
\mathcal{G}(k^2) = 4\pi^{4}\eta^{2}\delta(k) \, .
\end{equation}
It is obvious that such a model is an infrared-dominant model that
does not represent well the behavior in the large momentum region.
Substituting Eq.(13) into Eqs.(4) and (7), one can have the solution
with the nontrivial function $B(p^{2})$, which is related to the
dynamical quark mass, in the chiral limit
\begin{eqnarray}
A=2 \, , \qquad \qquad \quad \; & & B=\sqrt{\eta^{2}-4 p^2} \,  ,\\
C=-\frac{2}{\sqrt{(\eta^{2}-4p^{2}}} \, , & &
D=\frac{1}{2}\frac{\eta^{2}+8p^{2}}{\eta^{2}-4p^{2}} \, ,
\end{eqnarray}
for $p^{2}<\frac{\eta^{2}}{4}$, and
\begin{eqnarray}
A=\frac{1}{2}(1+\sqrt{1+\frac{2\eta^{2}}{p^2}}) \, , \, & & B=0 \,  ,\\
C=0 \, , \qquad  \qquad \qquad \quad    & &
D=\frac{p^{2}+\eta^{2}+p^{2}\sqrt{1+\frac{2\eta^{2}}{p^2}}}
{p^{2}-\eta^{2}+p^{2}\sqrt{1+\frac{2\eta^{2}}{p^2}}} \, ,
\end{eqnarray}
for $p^{2}\geq\frac{\eta^{2}}{4}$. With Eq.(10), the chiral
susceptibility in the infrared region can then be written as
\begin{equation}
\chi(p^2)=\frac{3}{4}\frac{\eta^{2}}{\eta^{2}-4p^{2}} \, .
\end{equation}
It should be noted that there exists a singularity at $p^{2} =
\frac{{\eta}^{2}}{4} $ in the chiral susceptibility in the chiral
limit. Such an evident pole in the susceptibility is an artifact of
the simplified nature of the Munczek-Nomirovsky model in the chiral
limit where the quark mass function becomes zero at finite
momentum\cite{Holl01}. A realistic description would not have such a
singular susceptibility. Making use of the Euclidean constituent
mass concept, one has $p^{2} = M_{c} ^{2} = \frac{{\eta} ^{2}}{8}$
in Eq.~(18) and it gives $\chi (p^2) = \frac{3}{2}$. With Eq.~(12),
one obtains the pion-nucleon sigma term in terms of the current
quark mass as
\begin{equation}
\sigma_{\pi N} = \frac{9}{2} m \,  .
\end{equation}
It is evident that, in the chiral limit approximation, the
pion-nucleon sigma term is estimated to be $9/2$ times the current
quark mass and independent of the strength of the infrared slavery
effect. If the current quark mass takes an empirical value about
$10$~MeV, we have apparently $\sigma_{\pi N} = 45$~MeV. It has been
shown that, if the potential representing the quark confinement can
be written as an exponential form $V_{c} \approx r^{Z}$, the
pion-nucleon sigma term can be given as $\sigma_{\pi N} =
\frac{9}{3-Z}m$\cite{Peng05}. Our present result is just the case of
linear confinement with $Z=1$. On the other hand, as a measure of
the uncertainty in such estimates of $\sigma_{\pi N}$, the
alternative estimate\cite{Craig05} of the constituent mass
$\tilde{M_{c}} = M (p^{2} = 0)$ gives $\tilde{M_{c}} =
\frac{\eta}{2}$, it results in $\chi(p^2 =0) = \frac{3}{4}$ and
$\sigma_{\pi N} = \frac{9}{4} m $. We consider this to overestimate
the current quark mass and underestimate $\sigma_{\pi N}$ because
the expected $p^2$ for a constituent quark in a nucleon should be
non-zero.

Beyond the chiral limit, we should calculate the Eqs.(4) and (7) at
finite quark mass with the Munczek-Nomirovsky model. The numerical
result of the sigma term at and beyond the chiral limit is shown in
Fig.~1. The Fig.~1 shows obviously that, if the current quark mass
is less than $15$~MeV, the effect beyond the chiral limit on the
pion-nucleon sigma term can be neglected. We take then the chiral
limit in the following discussions.
\begin{figure}[hbtp]
\begin{center}
\includegraphics[scale=1.5,angle=0]{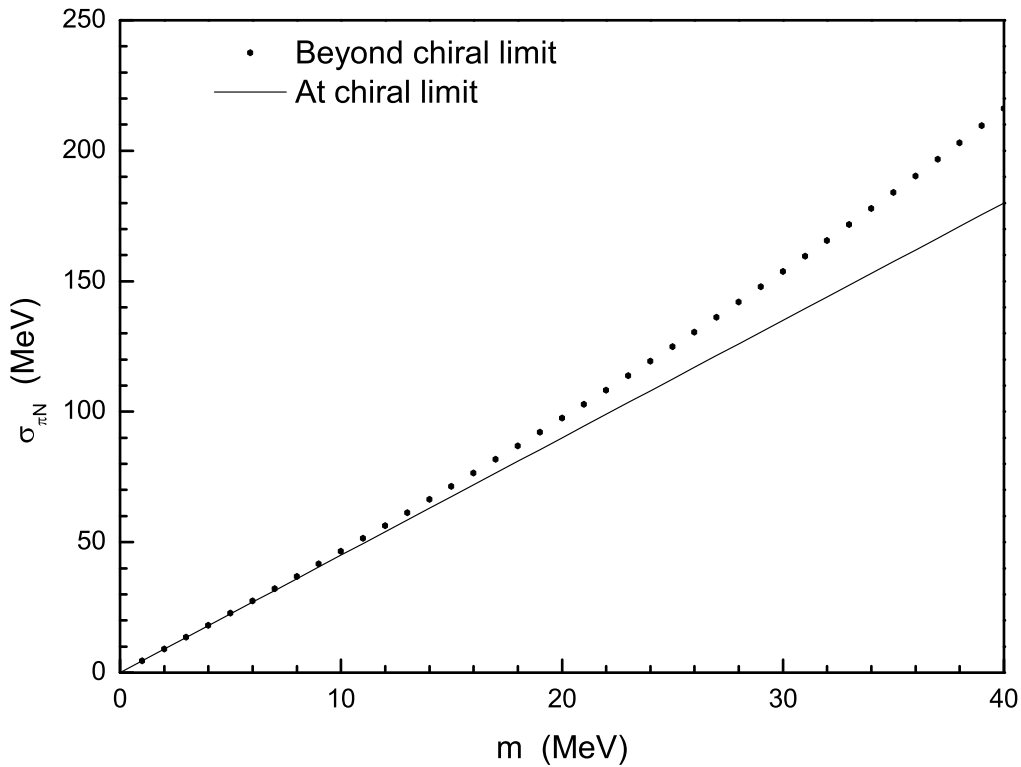}
\vspace*{0mm} \caption{\label{sigma-m} The current quark mass
dependence of the pion-nucleon sigma term at and beyond the chiral
limit.}
\end{center}
\end{figure}

From the relation between the sigma term and the constituent quark
mass we can infer that the sigma term is not a constant in nuclear
matter since the quark mass usually changes with the density. In the
GCM the quark propagator in-medium can be evaluated in a general
approach. It is then appropriate to study the sigma term in-medium
in the GCM.

The chemical potential is generally introduced as a Lagrangian
multiplier $e^{-\int d^{4}x\mu \bar{q}\gamma_{4}q}$ with the
partition function of GCM. With the general approach discussed
above, the quark equation and its derivation against the current
quark mass can be obtained. In the Munczek-Nomirovsky model (the
same form as in vacuum) at chiral limit, one can easily obtain the
functions $A$, $B$, $C$ and $D$. However, it is usually difficult to
numerically solve the the coupled D-S equation at finite chemical
potential with an general effective gluon propagator. In order to
avoid this difficulty, we have developed an approach for calculating
the chemical potential dependence of the dressed quark
propagator\cite{Chang01}. The main conclusion in the approach is
that the inverse of the full dressed quark propagator at finite
chemical potential can be obtained from the inverse of the dressed
quark propagator at zero $\mu$ by a replacement $p_{4}\rightarrow
p_{4}+i\mu$.

Taking the replacement $p_{4}\rightarrow p_{4}+i\mu$ and $p
\rightarrow \tilde{p} = (\vec{p}, p_{4}+i\mu)$, one can obtain the
functions $A$, $B$, $C$ and $D$ in Eq.~(10) at finite chemical
potential in the Munczek-Nomirovsky model easily. The quark chiral
susceptibility in Eq.(18) can then be generalized to that at finite
chemical potential $\mu > 0$ as
\begin{equation}
\chi( \mu )=\frac{3}{4}\frac{\eta^{2}}{\eta^{2}-4\tilde{p}^{2}} \, ,
\end{equation}
where $\tilde{p} ^{2} = {\vec{p}}^{2} + (p_{4} + i \mu)^{2}$.

Up to now, the quark constituent mass at finite chemical potential
has not yet been well defined. Analogous to that at finite
temperature, one has temporal mass identified by $M_{c} = M
(\vec{p}^{2} = 0, p_{4}^{2} = M_{c} ^{2})$. Another way is
identifying the mass to spatial mass by $M_{c} = M (\vec{p}^{2} =
M_{c} ^{2}, p_{4}^{2} = 0)$. Since the temporal mass takes a complex
number and so does the chiral susceptibility, we take the spatial
mass. As a consequence, we have the chemical potential dependent
pion-nucleon sigma term in Munczek-Nomirovsky model as
\begin{equation}
\sigma_{\pi N}(\mu) = \frac{9}{2}m
\frac{\eta^{2}}{\eta^{2}+4\mu^{2}} \, .
\end{equation}
It is apparent that, in the case of without medium (i.e., $\mu =0$),
$\sigma_{\pi N} (0) = \frac{9}{2} m$. Meanwhile, as the chemical
potential increases, the in-medium pion-nucleon sigma term
$\sigma_{\pi N}(\mu )$ decreases monotonously. It is again
remarkable that, if one takes the quark constituent mass for
chemical potential $\mu > 0$ as the value of the quark mass function
with all the components of momentum being zero, one has a factor
$\frac{1}{2}$ for the right hand side of Eq.~(21). This is parallel
to the situation at $\mu =0$ and the data can be regarded as the
lower limit at a certain chemical potential.

\section{Chiral Quark Condensate in Nuclear Matter}

As an application of the result of the chemical potential dependent
pion-nucleon sigma term, we discuss the chiral quark condensate in
hadron matter. It is well known that the in-medium chiral quark
condensate can be related to the pion-nucleon sigma term with the
model-independent relation\cite{CFG92}
\begin{equation}
\frac{<\bar{q}q>_{\rho}}{<\bar{q}q>_{vac}} = 1 -
\frac{1}{2|<\bar{q}q>_{vac}|}\frac{d\varepsilon}{dm} \, ,
\end{equation}
where $\rho$ is the density of the nuclear matter which can be fixed
by model calculation. $<\bar{q}q>_{\rho}$ is the quark condensate in
the nuclear matter with density $\rho$ and $<\bar{q}q>_{vac}$ is the
one in the vacuum which may depend on the current quark mass,
$\varepsilon$ is the energy density of the nuclear matter which can
be approximately written as\cite{CFG92}
\begin{equation}
\varepsilon=\rho M_{N} + \delta \varepsilon  \, ,
\end{equation}
where $M_{N}$ is the nucleon mass, and $\delta \varepsilon$ is the
contribution to energy density from the nucleon kinetic energy and
nucleon-nucleon interaction energy which is of higher order in
nucleon density and is empirically small at low density. Ignoring
the last term $\delta \varepsilon$ and the quark current mass
dependence of the density $\rho$, we have
\begin{equation}
\frac{<\bar{q}q>_{\rho}}{<\bar{q}q>_{vac}} = 1 - \frac{
\sigma_{\pi N} }{2 m \vert \langle \bar{q}q \rangle _{vac} \vert }
\rho  \, .
\end{equation}
Such a relation has been commonly used to study the medium density
effect on the chiral quark condensate with a constant $\sigma_{\pi
N}$ (see for example Refs.\cite{CFG92,Fuji95,Brock96}).

In the last section, we have obtained the chemical potential
dependent pion-nucleon sigma term in nuclear matter. To evaluate the
in-medium chiral quark condensate with Eq.~(24), we should transfer
the chemical potential dependence to the matter density dependence.
Then it is necessary to derive the relation between the nuclear
matter density and the quark chemical potential. At the mean field
level, the pressure density of the matter can be written as
\begin{equation}
\mathcal{P}[\mu]=Trln[G^{-1}]-\frac{1}{2}Tr[\Sigma G] \, .
\end{equation}
With such a pressure, one usually define the nuclear matter density
as\cite{Craig00}
\begin{equation}
\rho=\frac{\partial\mathcal{P}}{\partial \mu}
\end{equation}

As mentioned above, the inverse of the quark propagator and the
quark self-energy in the Munczek-Nomirovsky model for the
``Nambu-Goldstone" solution can be written explicitly by replacing
the $p$ with $ \tilde{p} = ( \vec{p}, p_{4} + i \mu )$ in the
following forms
\begin{eqnarray}
G^{-1}(p^{2})& = & [i\gamma\cdot p A_{1}(p^{2})+B_{1}(p^{2})
]\theta(\frac{\eta^{2}}{4}-p^{2}) \nonumber \\ & & + [ i\gamma\cdot
p A_{2}(p^{2})+B_{2}(p^{2})] [ 1-\theta(\frac{\eta^{2}}{4}-p^{2}) ]
\, ,
\end{eqnarray}
%and the ``Wigner" solution
%\begin{equation}
%G^{-1}(p^{2})=i\gamma\cdot p A_{2}(p^{2})+B_{2}(p^{2}) \, ,
%\end{equation}
\begin{eqnarray}
\Sigma(p^{2})& = & [i\gamma\cdot p (A_{1}(p^{2}) - 1) + B_{1}(p^{2})
]\theta(\frac{\eta^{2}}{4}-p^{2}) \nonumber \\ & & + [ i\gamma\cdot
p (A_{2}(p^{2}) - 1) + B_{2}(p^{2})] [
1-\theta(\frac{\eta^{2}}{4}-p^{2}) ] \, ,
\end{eqnarray}
where $A_{1}(p^{2})$, $B_{1}(p^{2})$ have the form of Eq.(14) and
$A_{2}(p^{2})$, $B_{2}(p^{2})$ relate to Eq.(16), and
$\theta(\frac{\eta^{2}}{4}-p^{2})=1$ for $p^{2}<\frac{\eta^{2}}{4}$,
$\theta(\frac{\eta^{2}}{4}-p^{2})=0$ for
$p^{2}\geq\frac{\eta^{2}}{4}$.

It should be noted that the Eq.~(26) with Eqs.~(25), (27) and (28)
is divergent due to the vector part of the quark propagator. To give
a finite value of nuclear matter density we expand the pressure
density with quark loop to second order. It is found that the second
order is zero at mean field level and the relation between the
nuclear matter density and the chemical potential $\mu$ can be
written as
\begin{equation}
\rho=\frac{2}{3\pi^{2}}\mu^{3}+4\frac{\partial}{\partial\mu}
\int\frac{d^{4}q}{(2\pi)^{4}}[A(\tilde{q}^{2})-1] \, .
\end{equation}

On the other hand, when applying the replacements $p \rightarrow
\tilde{p} = (\vec{p} , p_{4}+i\mu )$ and  $q \rightarrow \tilde{q} =
(\vec{q} , q_{4}+i\mu )$ in Eqs.~(26)-(29), one should do that not
only for the argument of functions $A_{1}, B_{1}, A_{2}, B_{2}$, but
also for that in the step function $\theta$. It is evident that the
boundary condition is arbitrary for the Heavyside step function
$\theta(\frac{\eta^{2}}{4}-\tilde{q}^{2})$, which cannot be defined
from any criterion\cite{Bender01}. Then we propose a choice as
\begin{equation}
Re[\frac{\eta^{2}}{4} - \tilde{q}^{2} ] > a\mu^{2} \,
\end{equation}
for the non-zero scalar function $B$, where $a$ is a parameter (the
choice $a=0$ has been taken in previous calculations with the
Munczek-Nomirovsky model (see for example Refs.\cite{MRS98,LGG03})).
Because the parameter $a$ determines the domains of different forms
of function $A$ and of the integration in Eq.~(29), it influences
the relation between the nuclear matter density and the chemical
potential. The numerical results of $\rho$ in terms of $\mu$ at
several parameters $a$ are showed in Fig.~2.
\begin{figure}[hbtp]
\begin{center}
\includegraphics[scale=1.5,angle=0]{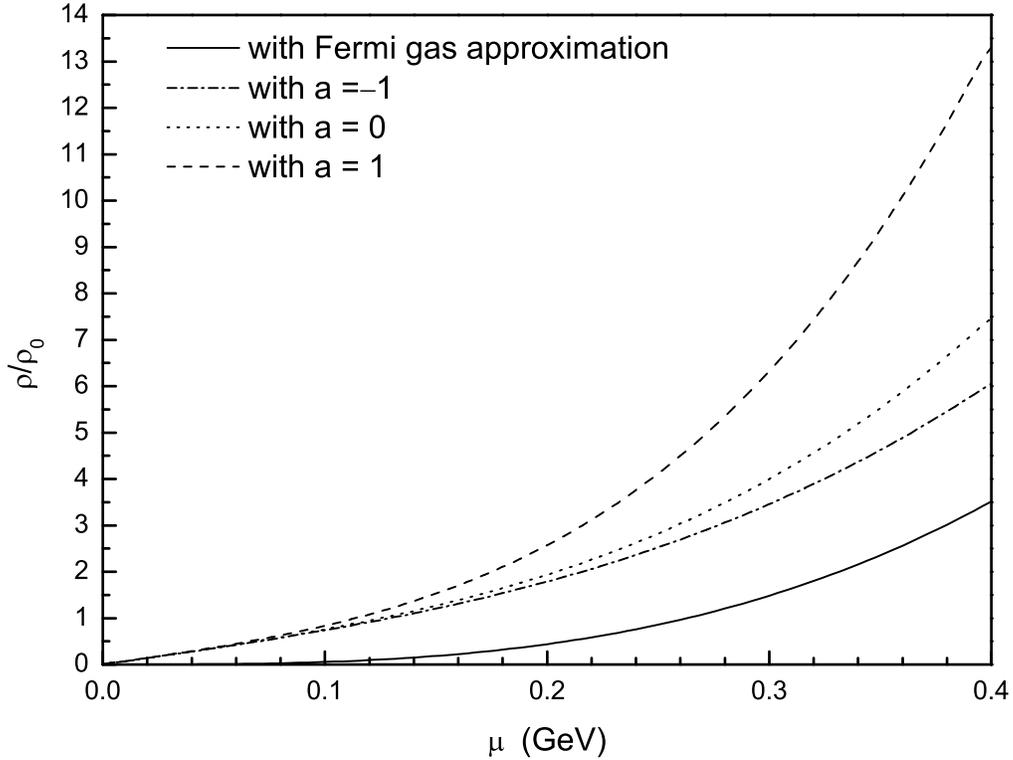}
\vspace*{0mm} \caption{\label{denst-chemp} Calculated relation
between the density and the chemical potential of nuclear matter
in several models.}
\end{center}
\end{figure}

Taking the relation between the nuclear matter density and the
chemical potential in Eq.~(29) into account, the chemical potential
dependence of the pion-nucleon sigma term in nuclear matter in
Eq.~(21) can be rewritten as a relation between the pion-nucleon
sigma term and the nuclear matter density. Basing on the result
shown in Fig.~2, that the density increases monotonously with the
chemical potential, one can recognize that the pion-nucleon sigma
term in nuclear matter decreases monotonously with the matter
density.

With the obtained relation between the pion-nucleon sigma term and
the nuclear matter density $\sigma_{\pi N} (\rho )$ being
substituted into Eq.~(24), we obtain the nuclear matter density
dependence of the chiral quark condensate in nuclear matter. The
results with current quark mass being taken as $m=10$~MeV and the
parameter $\eta$ being taken as $\eta = 1.04$~GeV, with which the
pion decay constant in free space ($93$~MeV) has been reproduced
well\cite{FT92}, are illustrated in Fig.~3. From Fig.~3, one can
recognize easily that, if the nuclear matter density dependence of
the pion-nucleon sigma is neglected, the chiral quark condensate
decreases linearly with the increase of the nuclear matter density
and reaches zero as the density is a little larger than 4 times the
normal nuclear matter density. As the nuclear matter density
dependence of the sigma term is taken into account, the decreasing
rate of the chiral quark condensate in nuclear matter is weakened
evidently. Meanwhile, the decrease maintains monotonous, so that the
``upturn" problem in some of the previous works does not emerge in
our present work. Comparing this result with those given in
Refs.\cite{Dru013,Lutz98}, one can infer that the inclusion of the
density dependence of sigma term is, in some sense, equivalent to
taking the higher order corrections or the effect of pion into
account appropriately. Furthermore, the degree of the reduction on
the decreasing rate depends on the boundary condition we proposed.
With the decrease of the boundary condition parameter $a$ from $1$
to $-1$, the decreasing rate gets obviously smaller. For instance,
the condensate reaches zero as $\rho \approx 5.5 \rho_{0}$ for
$a=1$, however the condensate vanishes as $\rho \approx 7.2
\rho_{0}$ for $a=-1$. Recalling Eq.~(30), one can know that such a
phenomenon means that the decreasing rate becomes smaller as the
domain for the scalar self-energy function $B \ne 0$ is enlarged. As
{\it vice versa}, it indicates that the boundary condition is a
manifestation of the interaction being taken into account.
\begin{figure}[hbtp]
\begin{center}
\includegraphics[scale=1.5,angle=0]{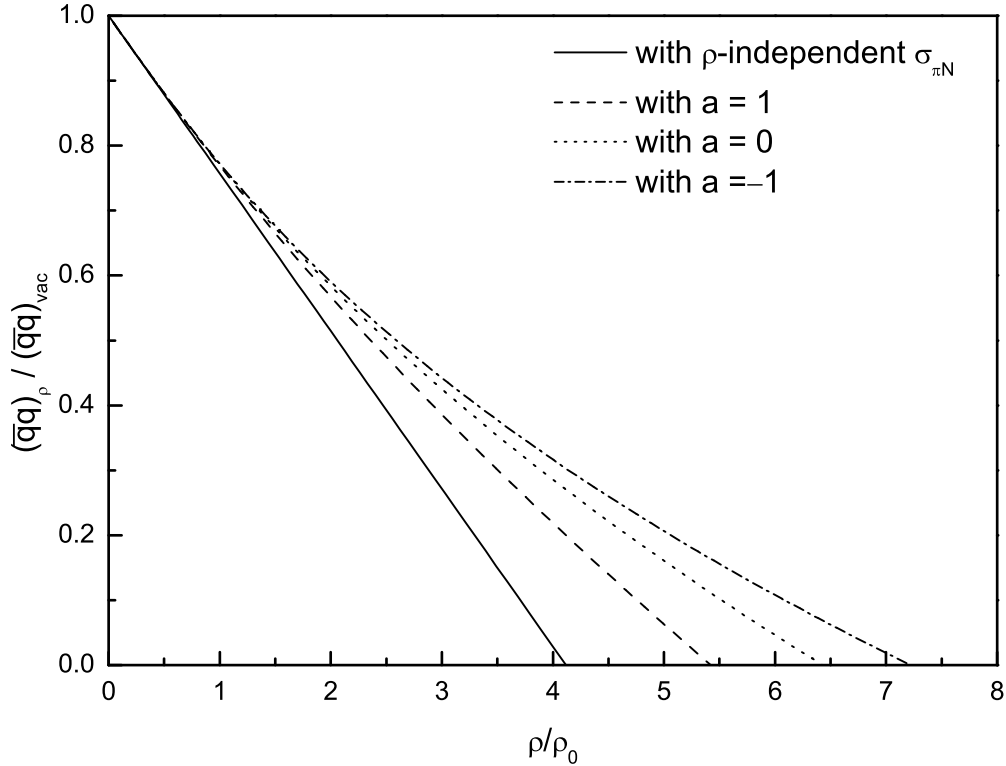}
\vspace*{0mm} \caption{\label{dpqcond} The calculated density
dependence of the chiral quark condensate with and without the
sigma term in-medium effect being taken into account. }
\end{center}
\end{figure}

\section{Summary and Remark}

In summary, we studied the pion-nucleon sigma term in vacuum and in
nuclear matter in the global color model of QCD with the effective
gluon propagator in Munczek-Nomirovsky model in this paper. It is
estimated that the pion-nucleon sigma term at chiral limit in the
vacuum is 9/2 times the current quark mass and the in-medium
pion-nucleon sigma term decreases with the nuclear matter density.
As an application of the obtained chemical potential (or hadron
matter density) dependence of the pion-nucleon sigma term, we take
it to study the in-medium chiral quark condensate. It shows that,
with the medium effect on the sigma term being taken into account,
the linear decreasing behavior is evidently shifted to nonlinear
with a much smaller descend rate. It indicates that such a variation
behavior of the pion-nucleon sigma term against the medium density
is reasonable and its effect on the chiral quark condensate in
nuclear matter is consistent with that takes the higher order
interaction into account.

Concerning our derivation of the pion-nucleon sigma term, we take
the Munczek-Nomirovsky model in the framework of global color model
of QCD and obtain an analytical expression in terms of the current
quark mass and the chemical potential. However, since the
Munczek-Nomirovsky model expresses the effective gluon propagator as
a $\delta$-function in momentum space, it can only represent well
the characteristic in infrared region but can not display the
behavior in ultraviolet region. Then the study with a more realistic
effective gluon propagator is necessary. Moreover, we take a nucleon
simply as three non-interacting quarks to evaluate the pion-nucleon
sigma term in the present work. The realistic nucleon is quark bound
state with complicated interactions. To obtain the pion-nucleon
sigma term more sophisticatedly needs to take into account the
interaction among the quarks. The related investigations are under
progress.

\bigskip

\bigskip

\begin{acknowledgments}
This work was supported by the National Natural Science Foundation
of China (NSFC) under contract Nos. 10425521, 10135030, 10075002,
10275002, 10435080, the Major State Basic Research Development
Program under contract No. G2000077400, the research foundation of
the Ministry of Education, China (MOEC), under contact No. 305001
and the Research Fund for the Doctoral Program of Higher Education
of China under grant No. 20040001010.
One of the authors (YXL) thanks also the support of the Foundation
for University Key Teacher by the MOEC. The authors are indebted
to Dr. Craig D. Roberts for his stimulating discussions.

\end{acknowledgments}

\newpage

\end{document}